%

\magnification \magstep1
\baselineskip 20pt
\def\pmb#1{\setbox0=\hbox{$#1$}%
  \kern-0.25em\copy0\kern-\wd0
  \kern.05em\copy0\kern-\wd0
  \kern-0.025em\raise.0433em\box0}
\def\bPsi{\pmb{\Psi}}

\def\prref{\par\noindent\hangindent=0.7cm\hangafter=1}

\def \cntl{\centerline}
\def \etal {{\it et al. }}
\def \dag {^\dagger}

\def \ie {{\it i.e.}, }
\def \eg {{\it e.g.}, }

\def \Nature {{Nature}}
\def \ApJ {{Astrophys. J.}}

\def \ARAA {{Ann. Rev. of Astron. \& Astrophys.}}
\def \ApJSup {{Astrophys. J. (Supp.)}}
\def \MNRAS  {{M.N.R.A.S.}}

\def \vjec {\vfill\eject}

\def\cL{{\cal L}}

\def \hmpc{h^{-1}Mpc}

\def\vr{{\bf r}}

\def \r0p{ r{_0^\prime}}

\def\br{{\bf r}}

\def\bd{{\bf d}}

\def\bm{{\bf m}}

\def \r0p{ r{_0^\prime}}

\def \br{{\bf r}}

\def\det{{\rm det}}
\def\m3{{Mark III}}

\def\today{\ifcase\month\or
  January\or February\or March\or April\or May\or June\or
  July\or August\or September\or October\or November\or
December\fi
  \space\number\day, \number\year}
%
%


%
%

%
%

\font\title=cmbx10 scaled 2000

\font\figtitle=cmbx10

\def\ss{\vskip 0 true cm}


\magnification \magstep1
\baselineskip 12pt
\parskip 4pt

\def\la{\langle} \def\ra{\rangle}
\def\etal{{\it et al.\ }} \def\rms{{\it rms\ }}
 \def\eg{{\it e.g.\ }} \def\ie{{\it i.e.\ }}

\def\lcdm{$\Lambda$CDM} \def\ocdm{OCDM}

\def\hmpc{\, h^{-1} {\rm Mpc}}
\def\ihmpc{\, h\, {\rm Mpc^{-1}}}  \def\3hmpc{\, ( h^{-1} {\rm Mpc})^3}
  \def\kmsmpc{\, {\rm km\,s^{-1} Mpc^{-1} }}
\def\h5{h_{50}}
\def\Pr{{\cal P}} 
\def\sbf{\sevenbf}
\def\srm{\sevenrm}
\def\ss{\scriptstyle}
\def\sslcdm{${\ss \Lambda}$CDM}
\def\secpush{\vskip0pt plus 2.0\baselineskip\penalty-250
	     \vskip0pt plus -2.0\baselineskip }
%


%

\vbox{\vskip 1truecm}

\cntl{\bf LARGE-SCALE POWER SPECTRUM FROM PECULIAR VELOCITIES}
\cntl{\bf VIA LIKELIHOOD ANALYSIS}

\bigskip
\cntl{\bf Saleem Zaroubi$^1$, Idit Zehavi$^2$, Avishai Dekel$^{1,2}$,}
\cntl{\bf Yehuda Hoffman$^2$ \& Tsafrir Kolatt$^{3,4}$}
\bigskip
\noindent
$^1${Astronomy Department and Center for Particle Astrophysics,
University of California, \hfill\break
 \indent Berkeley, CA 94720} \hfill\break
$^2${Racah Institute of Physics, The Hebrew University,
Jerusalem 91904, Israel} \hfill\break
$^3${Harvard-Smithsonian Center for Astrophysics, 60 Garden St.,
Cambridge MA 02138} \hfill\break
$^4${UCO/Lick Observatory, University of California, Santa Cruz,
CA 95064}

\bigskip\bigskip

\cntl{\bf ABSTRACT}

\baselineskip 12pt
\parskip 4pt
\rightskip=0.8 true cm \leftskip=0.8 true cm

The power spectrum (PS) of {\it mass} density fluctuations,
independent of ``biasing",
is estimated from the \m3 catalog of peculiar velocities
using Bayesian statistics. A parametric model is assumed for
the PS, and the free parameters are determined by maximizing the
probability of the model given the data.
The method has been tested using detailed mock catalogs.
It has been applied to
generalized CDM models with and without COBE normalization.

The robust result for all the models is a relatively high PS, with
$P(k) \Omega^{1.2} = (4.8\pm1.5)\times 10^3 \3hmpc$ at $k=0.1\ihmpc$.
An extrapolation to smaller scales using the different CDM models yields 
$\sigma_8 \Omega^{0.6} = 0.88 \pm 0.15$.
The peak is weakly constrained to the range $0.02\leq k\leq 0.06\ihmpc$. 
These results are consistent with a direct computation of the PS
(Kolatt \& Dekel 1996).
When compared to galaxy-density surveys, the implied values for
$\beta$ ($\equiv \Omega^{0.6}/b$) are of order unity to within 25\%.

The parameters of the COBE-normalized, flat CDM model are confined
by a 90\% likelihood contour of the sort
$\Omega\, {\h5}^\mu\, n^\nu = 0.8 \pm 0.2$,
where $\mu = 1.3$ and $\nu = 3.4,\ 2.0$ for models with and without
tensor fluctuations respectively.
For open CDM the powers are $\mu = 0.95$ and $\nu = 1.4$
(no tensor fluctuations). 
A $\Gamma$-shape model free of COBE normalization yields
only a weak constraint: $\Gamma = 0.4\pm 0.2$.

\bigskip\bigskip

\noindent{\it Subject headings:}
cosmology: theory --- cosmology: observation --- dark matter
--- galaxies: clustering --- galaxies: distances and redshifts
--- large scale structure of universe

\rightskip=0 true cm \leftskip=0 true cm

\vjec
\baselineskip 12pt
\parskip 4pt

\cntl{\bf 1. INTRODUCTION}

In the standard picture of cosmology, the structure on large scales
originated from small-amplitude initial density fluctuations that
were amplified by gravitational instability.
These initial fluctuations are assumed to be a Gaussian random field,
solely characterized by its power spectrum.
On large scales, the fluctuations are linear even at late times,
so that the power spectrum preserves its original shape.
This makes it a very useful statistics for large-scale structure.

The power spectra of galaxy density were derived for many different
samples, in two angular dimensions or in three dimensions from redshift
space.  Unfortunately, these power spectra correspond to objects that are not
necessarily unbiased tracers of the underlying mass distribution,
and it is the mass distribution that is directly related to theory
(\eg Dekel \& Rees 1987 for a review on ``galaxy biasing").
Clear evidence for this bias is provided by the fact that
galaxies of different types
are observed to cluster differently (\eg Dressler 1980).
It would therefore be naive to assume that any of the galaxy
power spectra directly reflects the mass PS.
Furthermore, in estimates of the galaxy PS from redshift surveys,
uncertainties also arise when correcting for redshift distortions
(Kaiser 1987, Zaroubi and Hoffman 1996). For these reasons, one wishes
to measure the
mass PS directly from dynamical data, bypassing the complex galaxy-biasing
issues and the need to correct for redshift distortions.
In principle, such dynamical information can be provided
by peculiar velocities, by gravitational lensing effects, or by
fluctuations in the cosmic microwave background (CMB).
In particular, the accumulating catalogs of galaxy peculiar velocities
enable a direct determination of the mass PS
under the natural assumption that the galaxies are
unbiased tracers of the large-scale, gravitationally-induced velocity field.

The PS is computed here from the \m3 catalog of peculiar velocities
(Willick \etal 1995 WI; 1996a WII; 1996b WIII), which
consists of more than 3000 galaxies. It was compiled from
several different data sets of spiral and elliptical/S0 galaxies with
distances inferred by the forward Tully-Fisher and $D_n\!-\!\sigma$ methods.
These data were re-calibrated and self-consistently put together as a
homogeneous catalog for velocity analysis.
The catalog provides radial peculiar velocities and inferred distances
with errors on the order of $17-21\%$ of the distance per galaxy,
sampled nonuniformly out to distances of $\sim 80\hmpc$ from the Local
Group.

The catalog exists in two versions that differ by the method of correction
for Malmquist bias.  In the standard ``G" version, the galaxies are first
heavily grouped into $\sim 1200$ objects ranging from isolated field
galaxies to rich clusters. The grouping reduces the non-linear noise
on small scales and the resulting Malmquist bias.
Then the data are systematically corrected for the Malmquist bias (see
Dekel 1994).
As a reference, we also use the ``S" version,
where a more straightforward systematic correction is applied to the
field galaxies as singles, at the expense of larger noise on small
scales.

These data allow a reasonable recovery of the dynamical fields with
$\sim\!12\hmpc$ smoothing in a sphere of radius $\sim\!60\hmpc$ about
the Local Group, extending to $\sim\!80\hmpc$ in certain regions.
The POTENT method (Bertschinger \& Dekel 1989;
Dekel, Bertschinger \& Faber 1990; Dekel 1994, 1997)
attempts a recovery of the underlying density field with fixed Gaussian
smoothing within this volume.
In an associated paper, Kolatt and Dekel (1996, KD) have computed
the mass PS from the smoothed density field recovered by POTENT from \m3.
The limitations of the data introduce severe systematic errors, that were
modeled via Monte-Carlo mock catalogs and then used to correct the
measured PS.
Since the KD results naturally involve uncertainties, an 
independent estimate of the PS, using a very different method, is useful. 

Our purpose is to estimate the mass PS
directly from the peculiar velocities of the \m3 catalog, by means of a
likelihood analysis.
The non-local nature of the peculiar velocities, \ie being influenced
by the mass distribution in a whole neighborhood, allows
one to probe scales somewhat larger than those probed by the density field.
For example, the effect of a bulk velocity across the entire volume is
not evident if only the density field is considered.
For a similar reason, the velocity field is expected to obey linear
theory better than the density field smoothed on a comparable scale, and
to closer resemble
a Gaussian field. Our approach here does not involve any explicit
window function, weighting or smoothing,
nor does it require artificial binning of the PS.  In addition, it 
automatically underweights noisy, unreliable data. 

The data analyzed here are especially suited for Bayesian analysis.
The sparse and inhomogeneous sampling of a random Gaussian field with
Gaussian errors yields a multivariate Gaussian data set.
The corresponding {\it posterior} probability distribution function (PDF) is
a multivariate
Gaussian that is completely determined by the assumed PS and the assumed
covariance matrix of errors. Under these conditions one can write the joint PDF
of the model PS and the underlying velocity or density field,
and then simultaneously estimate the PS model parameters and recover the
``Wiener filter" solution of the fields (Zaroubi \etal 1995).
In an associated paper (Zaroubi, Hoffman \& Dekel 1996), we present the
high-resolution fields recovered from this same data set using the PS
derived here.

To apply our method,
the simplifying assumptions that have to be made are (a) that
the peculiar velocities are drawn from a Gaussian field,
(b) that their correlation function can be derived from
the density PS using linear theory, and (c) that the errors are Gaussian
and accurately estimated.  
The need to assume a parametric functional form for the PS
is also a limitation;
one can try to achieve flexibility by using a large number of parameters
and a variety of functional forms,
but at the risk of making the likelihood analysis unstable in some
cases (\S 5).

The method is described in \S 2, where the relation between the PS
and the velocity correlation functions is specified, and
the likelihood algorithm for computing the PS is described.
The method is tested using a mock catalog in \S 3.
The resultant power spectra are presented in \S 4,
as derived from the \m3 data alone, and for generalized
CDM models imposing COBE normalization.
The associated constraints on the cosmological parameters are analyzed.
Our conclusions are summarized and discussed in \S 5.

\bigskip\secpush
\cntl{\bf 2. METHOD}

\smallskip
\cntl{\it 2.1. Velocity Correlations}

The computation of the matter power spectrum from the peculiar velocity data
by means of likelihood analysis requires a relation between the
velocity correlation function and the power spectrum.
Define the two-point velocity correlation ($3\times 3$) tensor by the average
over all pairs of points $\vr_i$ and $\vr_j$ that are separated by
$\vr=\vr_j-\vr_i$,
$$
\Psi_{\mu\nu} (\vr) \equiv \la v_\mu (\vr_i) v_\nu (\vr_j) \ra \,,
\eqno (1)
$$
where $v_\mu(\vr_i)$ is the $\mu$ component of the peculiar velocity
at $\vr_i$.
In linear theory, it can be expressed in terms of two scalar functions
of $r=\vert \vr \vert$ (G\'orski 1988), parallel and perpendicular to
the separation $\vr$,
$$
\Psi_{\mu\nu}(\vr)
= \Psi_{\perp}(r) \delta_{\mu\nu} +
[\Psi_{\Vert}(r) - \Psi_{\perp}(r)] \hat \br_\mu \hat \br_\nu \,.
\eqno (2)
$$
The spectral representation of these radial correlation functions is
$$
\Psi_{\perp,\Vert}(r)= {H_0^2 f^2(\Omega)\over 2 \pi^2}
\int_0^\infty P(k)\, K_{\perp,\Vert}(kr)\, dk \,,
\eqno (3)
$$
where $K_{\perp}(x) = j_1(x)/ x$ and
$K_{\Vert}(x) = j_0-2{j_1(x)/ x}$,
with $j_l(x)$ the spherical Bessel function of order {\it l}.
The cosmological $\Omega$ dependence enters as usual in linear theory
via $f(\Omega)\approx \Omega^{0.6}$, and $H_0$ is the Hubble constant.
A parametric functional form of $P(k)$ thus translates to a parametric
form of $\Psi_{\mu\nu}$.

\medskip\secpush
\cntl{\it 2.2. Likelihood Analysis}

Let $\bm$ be the vector of model parameters and $\bd$ the vector of $N$
data points. Then Bayes theorem states that the {\it posterior}
probability density of a model given the data is
$$
\Pr (\bm \vert \bd ) = {\Pr(\bm) \Pr(\bd|\bm) \over \Pr(\bd)} \,.
\eqno (4)
$$
The denominator is merely a normalization constant.
The probability density of the model parameters, $\Pr(\bm)$, is unknown,
and in the absence of any other information we assume it is uniform
within a certain range.
The conditional probability of the data given the model,
$\Pr(\bd|\bm)$, is the likelihood function, ${\cal L}(\bd|\bm)$.
The objective in this approach, which is finding the set of parameters
that maximizes the probability of the model given the data, is thus
equivalent to maximizing the likelihood of the data given the model
(Kaiser 1988; see also Jaffe \& Kaiser 1994 for a first application 
to the Lauer \& Postman data). 

The Bayesian analysis measures the relative likelihood of different models.
An absolute frequentist measure of goodness of fit
could be provided by the Chi-square per degree of freedom,
which we use as a check on the best parameters obtained by the
likelihood analysis.

Assuming that the velocities form a Gaussian random field,
the two-point velocity correlation tensor $\bPsi$ fully characterizes the
statistics of the velocity field.
Define the radial-velocity correlation ($N\times N$) matrix $U_{ij}$ by
$U_{ij} = \hat \br_i\dag\, \bPsi \,\hat\br_j$,
where $i$ and $j$ refer to the data points.
Let the inferred radial peculiar velocity at $\vr_i$ be $u_i$,
with the corresponding error $\epsilon_i$ also assumed to be a
Gaussian random variable.  The observed correlation matrix is then
$\tilde U_{ij} = U_{ij} + \epsilon_i^2 \delta_{ij}$, and
the likelihood of the $N$ data points is
$$
{\cal L} = [ (2\pi)^N \det(\tilde U_{ij})]^{-1/2}
  \exp\left( -{1\over 2}\sum_{i,j}^N {u_i \tilde U_{ij}^{-1} u_j}\right)\,.
\eqno (5)
$$

Given that the correlation matrix, $\tilde U_{ij}$, is symmetric and
positive definite, we can use the Cholesky decomposition method
(\eg Press \etal 1992) for computing the likelihood function (Eq.~5).
The significant contribution of the errors to the diagonal terms
makes the matrix especially well suited  for decomposition.
The calculation for a given choice of parameters and
$N\sim 1200$ data points takes a few minutes on a Dec-Alpha workstation
(of SpecFP92 $\sim 150$).

The likelihood function of Equation 5 is the posterior PDF
of the parameters $\bm$. It is a $\chi^2$ distribution (with $N$
degrees of freedom) with respect to the $N$ data points,
but it is not necessarily a $\chi^2$ distribution
with respect to the parameters.
Therefore, the task of assigning accurate confidence levels to the
parameters requires elaborate integrations over the volume encompassed
by the equal-likelihood surfaces in parameter space.
In the present paper we limit ourselves to a rough
estimate of confidence levels by crudely approximating
$-2 {\ln \cL}$ as a $\chi^2$ distribution in parameter space.

An important feature of this method, which distinguishes it from    
other methods (\eg KD), is that the formal likelihood errors        
include both the distance measurement errors and the cosmic scatter 
due to the finite discrete sampling.                                

Note that the grouping in the \m3 catalog also serves as a mean of smoothing 
over nonlinear velocities.  The PS on large-scales should not be much 
affected by this grouping because when group galaxies enter the 
likelihood analysis as individuals, they enter with low weights 
compared to the weight of a whole group.
For un-grouped galaxies, the noise term $\epsilon_i^2\delta_{ij}$ 
is larger (relative to the signal $U_{ij}$)
in the observed correlation matrix $\tilde U_{ij}$ of Equation 5.

Note also that the quantity that can be derived from peculiar velocity data via
the linear approximation is $f^2(\Omega)\, P(k)$, where
$P(k)$ is the mass density PS (see Eq.~3).

\bigskip\secpush
\cntl{\bf 3. TESTING THE METHOD}

Careful testing of the method with realistic mock catalogs is essential
in view of the large distance errors, the sparse and non-uniform sampling,
the bias-correction procedures, and the possible non-linear and
non-Gaussian effects.

The mock Mark III catalogs are described in Kolatt \etal (1996).
They are based on simulations whose
initial conditions were extracted from a reconstruction
of the smoothed real-space density field from the IRAS 1.2Jy redshift
survey, taken back into the linear regime.
Small-scale perturbations were added by means of constrained
random realizations. The system was then evolved forward
in time using an N-body simulation assuming $\Omega=1$,
and stopped at two alternative times,
when the \rms density fluctuation in a top-hat sphere of radius $8\hmpc$
reached $\sigma_8=0.7$ and later when $\sigma_8=1.12$.

 The ``galaxies" in the simulation were identified via a
linear biasing scheme (b=1.35), and they were divided into `spirals' and
`ellipticals' according to Dressler's morphology-density relation.
The galaxies were assigned TF quantities (internal velocities and absolute
magnitudes) that were Gaussianly scattered about an assumed TF relation,
and were then ``observed" following the selection criteria of the actual
data sets that compose the \m3 catalog.
The mock catalogs were grouped and corrected for biases just like the
real data, producing analogous G and S mock catalogs.

\bigskip

\def\myfig#1#2#3#4{
  \midinsert
     \vskip #2 true cm \vskip 2.3 true cm
     #3
     \vskip 5.6 true cm
     {\baselineskip 11pt
       \rightskip=-0.1 true cm \leftskip=0.2 true cm
       \par\noindent{\figtitle}\hskip -0.05 true cm  #4 \par
       \rightskip=0.1 true cm \leftskip=0.1 true cm
     }
     \vskip -0.4 true cm
  \endinsert
}
\myfig {1} {-0.4}
{\includegraphics{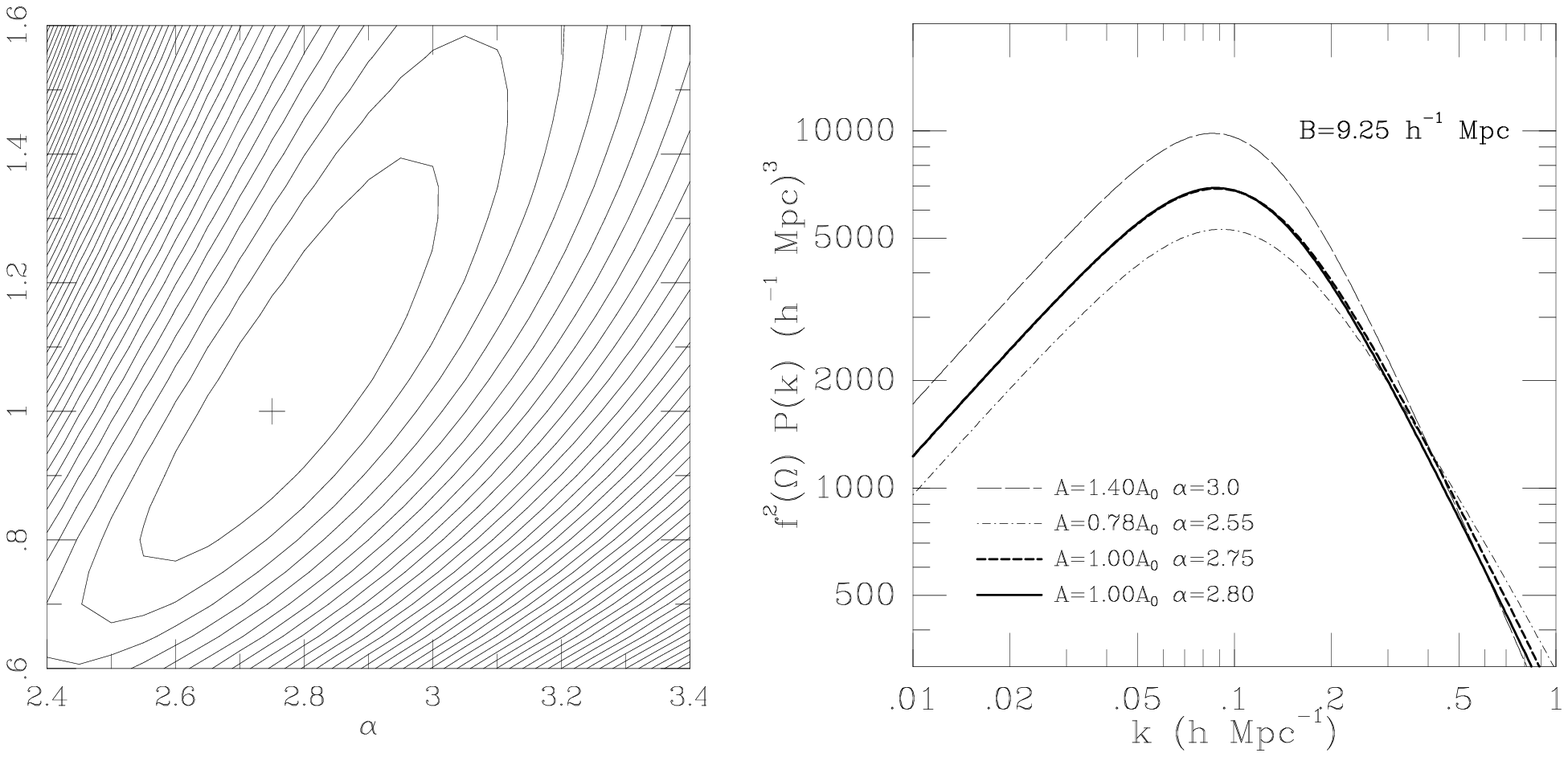}}
{{\sbf Figure 1a:} {\srm Contour map of ${\ss \ln}$-likelihood
in the ${\ss \alpha-A}$ plane for a mock catalog
based on the parametric model of Equation 6 with
${\ss B=9.25\hmpc}$. Contour spacing is ${\ss \Delta[\ln \cL] =-1}$.
The best-fit point is marked.}
\vskip 0 true cm \noindent
{\sbf Figure 1b:} {\srm The true power spectrum of the
simulation (heavy solid),
compared with the best-fit solution (heavy dashed), and two power
spectra whose parameters lie on the innermost closed contour of Figure 1a.}
\vskip 0.5 true cm
}

The true PS of the mass in the simulation is well approximated by the
functional form
$$
P(k) = {A_0\, k \over 1 + (B\, k)^\alpha} \,,
\eqno (6)
$$
with $A_0=4.68$ and $12.28 \times 10^4 (\hmpc)^4$,
$B=8.3$ and $9.25 \hmpc$, and $\alpha = 3.2$ and $2.8$,
for the $\sigma_8=0.7$ and $1.12$ cases respectively.

The likelihood analysis was applied to the mock catalogs using the
parametric functional form of Equation 6 as a prior.
Only two parameters were allowed to vary at a time while the third
parameter was kept fixed.
Figure 1a shows, for example, a contour map of $\ln \cL$ (relative to
the maximum-likelihood peak) for one of the $\sigma_8=1.12$, G
mock catalogs, spanning the $\alpha-A$ plane with $B=9.25\hmpc$.
The contours are separated by $\Delta[\ln \cL] = -1$.
Maximum likelihood is obtained at $A=A_0$ and $\alpha=2.75$
(compared to $2.80$).
Assuming a $\chi^2$ distribution with two degrees of freedom, the
$90\%$ confidence limit of the likelihood around the best-fit parameters
is at ${\ln \cL}\approx -2.3$.
We conclude that the recovery method works well. 

Figure 1b shows the recovered PS in comparison with the true PS
of the simulation.  They almost coincide over the whole range of scales,
showing slight deviations only on very small scales.
To illustrate the level of uncertainty, we plot for comparison
two other power spectra that were obtained with parameter pairs that lie
on the innermost contour
about the maximum in Figure 1a.
It shows that the amplitude near the peak can be off by about 25\%,
and that the recovery becomes more robust at moderately smaller scales.
The success of the recovery is similar when the other pairs of parameters
are allowed to vary, and also when allowing all three parameters to
vary at the same time.

Similar success was achieved when the method was applied to the
S mock catalogs. In what follows we focus on the results from
the G \m3 catalog, and refer to the S catalog as an indication for robustness.

\bigskip\secpush
\cntl{\bf 4. RESULTS}

\smallskip
\cntl{\it 4.1. The $\Gamma$ Model}

We first recover the PS from the velocity data alone,
independent of COBE normalization.
We use as a parametric prior the so-called $\Gamma$ model
(\eg Efstathiou, Bond and White 1992),
$$
P(k)= A\, k\, T^2(k), \quad
T(k) = \Bigl( 1 + [ ak/\Gamma + (bk/\Gamma)^{3/2} + (ck/\Gamma)^2 ] ^{\nu}
\Bigr)^{-1/\nu}\,,
\eqno (7)
$$
with $a=6.4\hmpc$, $b=3.0\hmpc$, $c=1.7\hmpc$ and $\nu = 1.13$.
The free parameters to be determined by the likelihood analysis are
the normalization factor $A$ and the $\Gamma$ parameter.
In the context of the CDM cosmological model, $\Gamma$ has a specific
cosmological interpretation, $\Gamma=\Omega h$. Here, however, independently
of CDM,  Equation 7 serves as a generic function with logarithmic slopes $n=1$
and $-3$ on large and small scales respectively, and with a turnover at
some intermediate wavenumber that is determined by the single shape
parameter $\Gamma$.

\def\myfig#1#2#3#4{
  \midinsert
     \vskip #2 true cm \vskip 2.3 true cm
     #3
     \vskip 5.6 true cm
     {\baselineskip 11pt
       \rightskip=-0.1 true cm \leftskip=0.2 true cm
       \par\noindent{\figtitle}\hskip -0.04 true cm #4 \par
       \rightskip=0.1 true cm \leftskip=0.1 true cm
     }
     \vskip -0.4 true cm
  \endinsert
}
\myfig {1} {-0.3}
{\includegraphics{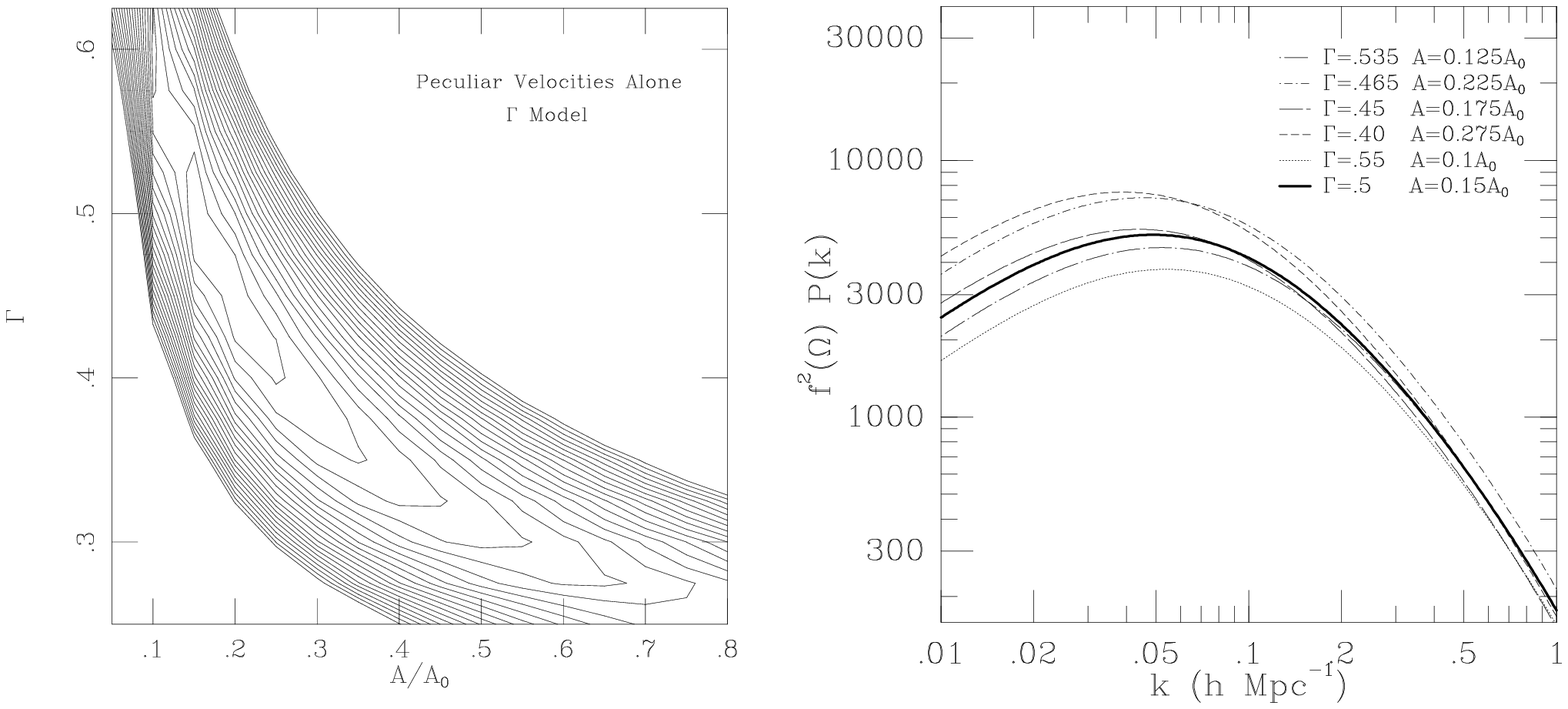}}
{{\sbf Figure 2a:} {\srm Contour map of ${\ss \ln}$-likelihood
for the ${\ss \Gamma}$ model.
Contour spacing is ${\ss \Delta[\ln\cL]=-1}$. ${\ss A}$ in units of
${\ss A_0 = 2.0\times 10^6 (\hmpc)^4}$.}
\vskip 0 true cm \noindent
{\sbf Figure 2b:} {\srm The most likely ${\ss \Gamma}$-model power spectrum
(solid), and five other models whose parameters lie on the innermost contour
of Figure 2a.}
\vskip 0.35 true cm  }

Figure 2a shows the contour map of $\ln \cL$ in the $A-\Gamma$
plane. The maximum likelihood values are $\Gamma=0.5 \pm 0.15$ and
$A=3.0\times 10^5 (\hmpc)^4$. 
The corresponding value of $\sigma_8\Omega^{0.6}$ is $0.85 \pm 0.1$.
The error-bars quoted here (and throughout the paper) are the
$90\%$ confidence limits.
Figure 2b shows the best-fit PS (solid).
To illustrate the uncertainty in the PS we also show
the power spectra of five other parameter pairs
that lie on the innermost likelihood contour about the best fit
(roughly the $65\%$ confidence limit).
We find the estimate of $\sigma_8 \Omega^{0.6}$ to be robust to the grouping
of the data. On the other hand, the estimate of $\Gamma$ without
imposing COBE normalization is sensitive to the grouping;
for the S catalog we obtain $\Gamma=0.3\pm 0.1$.

\medskip
\medskip\secpush
\cntl{\it 4.2. COBE-Normalized CDM Models}

We now restrict our attention to the generalized family of CDM cosmological
models, allowing variations in the cosmological parameters $\Omega$, $\Lambda$
and $h$, as well as the large-scale PS slope $n$ and the contribution
of tensor fluctuations.  Furthermore,
we now impose the normalization implied by the
four-year COBE DMR data (Hinshaw \etal 1996) 
as an additional external constraint.
The general form of the PS in these models is
$$
P(k) = A_{\fiverm COBE}(n,\Omega,\Lambda)\,
T^2(\Omega,\Omega_B,h; k)\, k^n\,,
\eqno (8)
$$
where we adopt the CDM transfer function proposed by Sugiyama (1995,
a slight modification of Bardeen \etal 1986): 
$$
T(k) = {\ln\left(1+2.34q) \right) \over 2.34q}
\left[1+3.89q+(16.1q)^2+(5.46q)^3+(6.71q)^4\right]^{-1/4}\,,
\eqno(9a)
$$
$$
q=k \left[ \Omega h\,
	   \exp (-\Omega_b -\h5^{1/2} \Omega_b/\Omega)\,
	   (\ihmpc)  \right]^{-1}
\eqno(9b)
$$
($\h5 \equiv H_0/50\kmsmpc = 2h$).
The parameters are varied, two at a time, such that they span the range
of currently popular CDM models, including
Tilted-$\Lambda$ CDM (flat: $\Omega+\Lambda =1$, $\Omega\leq 1$, $n\leq 1$)
and Tilted-Open CDM ($\Lambda=0$, $\Omega\leq 1$, $n\leq 1$).
We allow the possibility of nonzero tensor fluctuations, $T/S = 7 (1-n)$,
where the ratio is of quadrupole moments ($C_2$) of tensor and
scalar modes in the expansion of angular temperature fluctuations.
(\eg Turner 1993; Crittenden \etal 1993).
In all cases, the baryonic density is assumed to be $\Omega_b =
0.024 h^{-2}$ (\eg Tytler \etal 1996).  

The COBE normalization for each model has been calculated by various authors
(G\'orski \etal 1995; Sugiyama 1995; White \& Bunn 1995),
using different Boltzmann codes, different statistical analyses, and
sometimes even different temperature maps.
We have arbitrarily adopted Sugiyama's normalization as a backbone,
and for models not studied by him we use the other results after
matching them to Sugiyama's, using the models that they have investigated in
common.

In particular,
the COBE normalization is modeled by $A_{\fiverm COBE}= A_1(\Omega)\, A_2(n)$.
For Tilted-$\Lambda$ CDM models we use the fits:
$$
\eqalignno{
{\log A_1(\Omega)} = 7.83 - 8.33\Omega & + 21.31\Omega^2 - 29.67\Omega^3
+ 10.65\Omega^4 + 15.42\Omega^5 - \cr
& 6.04\Omega^6 - 13.97\Omega^7 + 8.61\Omega^8\,,
& (10a) \cr}
$$
$$
{\log A_2(n)} =\cases{-2.78+2.78n & $T/S=0$ \cr
		      -4.54+4.54n & $T/S\not=0$}\,.
\eqno(10b)
$$
These fits are for $h=0.5$, but the $h$ dependence in the range of interest
is weak, and we ignore it here.
For Tilted-Open CDM model with $T/S=0$ the fit is:
$$
{\log A_1(\Omega)} = 5.70 + 1.68\Omega - 4.53\Omega^2 + 7.57\Omega^3 -
7.53\Omega^4 + 3.15\Omega^5 - 0.23\Omega^6\,,
\eqno(11a)
$$
$$
{\log A_2(n)} = -2.71+2.71n \,.
\eqno (11b)
$$

\def\myfig#1#2#3#4{
  \midinsert
     \vskip #2 true cm \vskip 2.3 true cm
     #3
     \vskip 5.2 true cm
     {\baselineskip 11pt
       \rightskip=0.1 true cm \leftskip=0.1 true cm
       \par\noindent{\figtitle}\hskip -0.04 true cm  #4 \par
       \rightskip=0 true cm \leftskip=0 true cm
     }
     \vskip -0.6 true cm
  \endinsert
}
\myfig {1} {0.3}
{\includegraphics{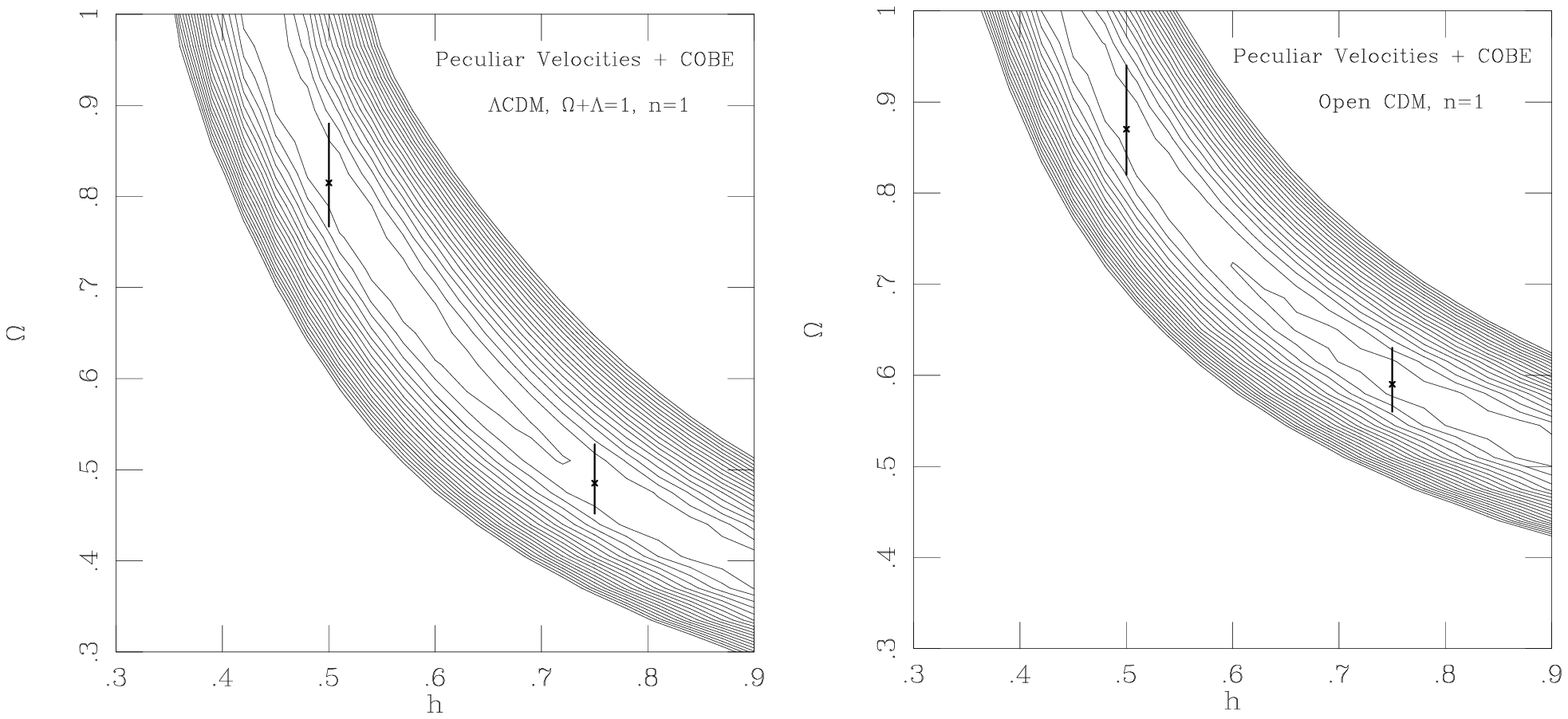}}
{{\sbf Figure 3:} {\srm Contour map of ${\ss \ln}$-likelihood in the
${\ss h-\Omega}$ plane for the \sslcdm\ (a) and \ocdm\ (b) models with
${\ss n=1}$. ${\ss \Delta[\ln \cL] = -1}$.
Shown are the most likely values of ${\ss \Omega}$ for two
fixed values of ${\ss h}$, and the
error bars corresponding to ${\ss 90 \%}$ confidence level.}
\vskip 0.2 true cm
}

\medskip\noindent\secpush\noindent
{\it 4.2.1 Scale-Invariant Models}

Figure 3 shows the likelihood contour map, in the $h-\Omega$
plane, for the flat CDM (\lcdm) and Open CDM (\ocdm) families of
models with $n=1$ (four-year COBE normalization by Sugiyama).
It is clear from the elongated contour pattern that $\Omega$ and $h$ are not
constrained very effectively independently of each other.
It is a degenerate combination of the two parameters
that is being tightly determined by the elongated ridge of high likelihood.
The constraints in the range plotted can be approximated (eye-ball
fit) by the functional form
$$
\eqalignno{
\Omega\, {\h5}^{1.3} =  0.83\pm 0.09\,, \, \qquad & \Lambda CDM \,, & (12a) \cr
\Omega\, {\h5}^{0.95} = 0.88\pm 0.07\,, \qquad  & OCDM \,. & (12b) \cr}
$$
Also marked in Figure 3 are the most likely values of $\Omega$ for fixed
given values of $h=0.5$ and $h=0.75$, and the corresponding one-dimensional
error bars.
The apparent weak preferences for high $\Omega$ (\lcdm) or high $h$ (\ocdm)
along the ridges of high likelihood is insignificant. In fact, it is
not robust to changes between the G and S catalogs.

\def\myfig#1#2#3#4{
  \midinsert
     \vskip #2 true cm \vskip 2.3 true cm
     #3
     \vskip 2.5 true cm
     {\baselineskip 11pt
       \rightskip=0.1 true cm \leftskip=0.1 true cm
       \par\noindent{\figtitle}\hskip -0.03 true cm #4 \par
       \rightskip=0 true cm \leftskip=0 true cm
     }
     \vskip -0.1 true cm
  \endinsert
}
\myfig {1} {11.4}
{\includegraphics{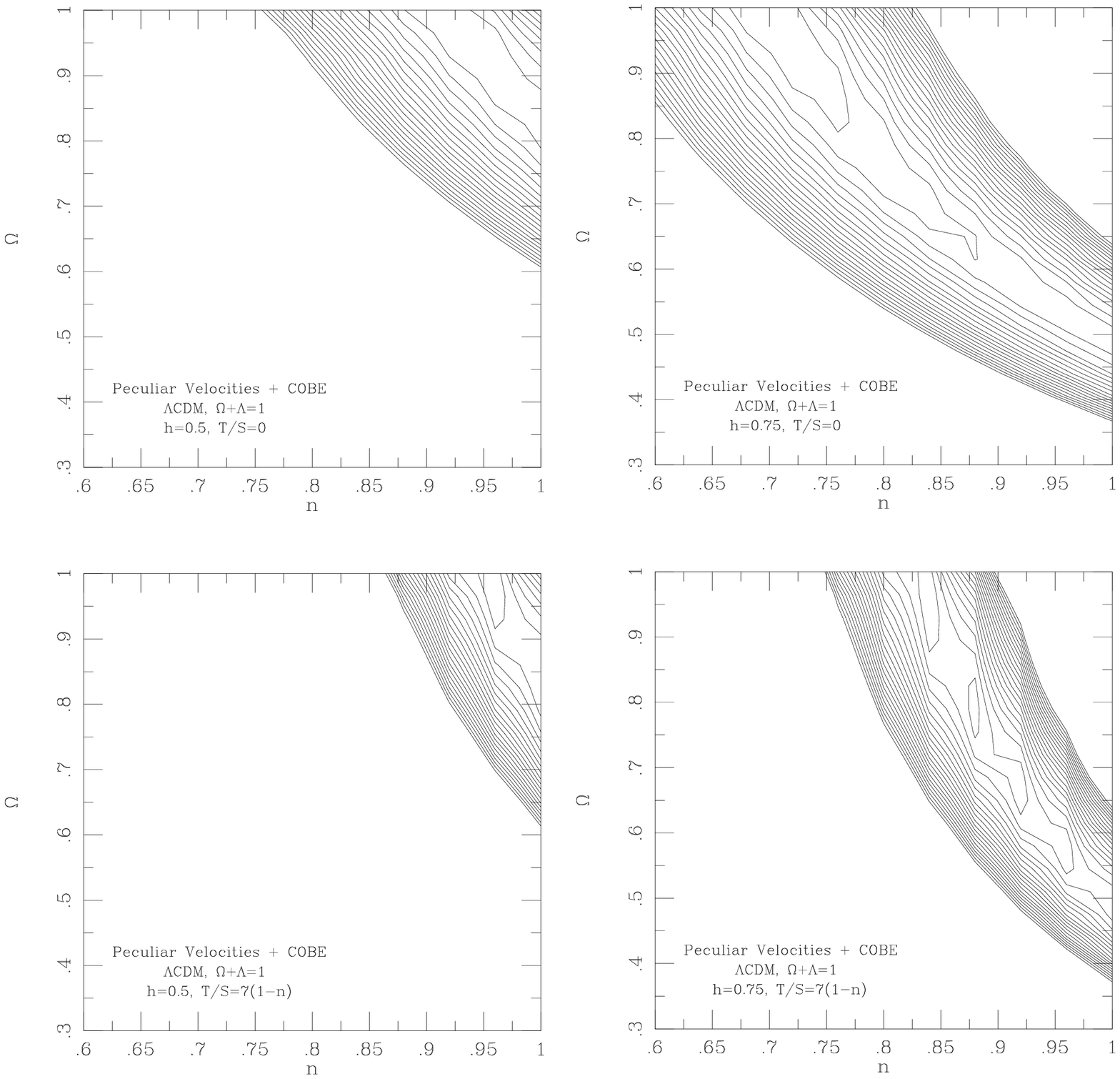}}
{\noindent
{\sbf Figure 4:} {\srm Contours of ${\ss \ln}$-likelihood in the
${\ss n-\Omega}$ plane, calculated with ${\ss h=0.5}$ and ${\ss 0.75}$
for tilted-\sslcdm\ models, with and without tensor component.
${\ss \Delta[\ln \cL] = -1}$.}
\vskip 0.3 true cm
}

\bigskip\noindent\secpush\noindent
{\it 4.2.2 Tilted Models}

Figure 4 shows the likelihood in the $n-\Omega$ plane
for the tilted-\lcdm\ family of models.
The computation is done for fixed values of $h$ at $h=0.5$ or $0.75$,
with or without tensor fluctuations.
COBE normalization is by White and Bunn (1995).
The elongated ridge of high likelihood can now be described by
$$
\eqalignno{
\Omega\, {\h5}^{1.3}\, n^{2.0} = 0.83 \pm 0.12\,, \qquad & T/S=0\,, & (13a) \cr
\Omega\, {\h5}^{1.3}\, n^{3.4} = 0.83 \pm 0.12\,, \qquad & T/S=7(1-n)\,. &
(13b) \cr}
$$
The trends along the ridges are, again, not robust to changing the
data from G to S.

For a fixed $\Omega$, this relation can be understood qualitatively as
follows:
the normalization by COBE fixes the amplitude at small wavenumbers,
$k\sim 0.001$, and the velocity data constrain the amplitude at $k\sim 0.1$.
The wavenumber corresponding to the peak of the PS is roughly proportional
to $\Omega h$.
Therefore, if a good fit is obtained with certain values of $h$ and $n$,
a similarly good fit can be obtained with higher $h$ and lower $n$,
or vice versa.
The presence of tensor fluctuations lowers the amplitude imposed by COBE
at small wavenumbers, and thus weakens the requirement for a tilt in $n$.

\def\myfig#1#2#3#4{
  \midinsert
     \vskip #2 true cm \vskip 2.3 true cm
     #3
     \vskip 5.3 true cm
     {\baselineskip 11pt
       \rightskip=0.1 true cm \leftskip=0.1 true cm
       \par\noindent{\figtitle}\hskip -0.03 true cm  #4 \par
       \rightskip=0 true cm \leftskip=0 true cm
     }
     \vskip -0.4 true cm
  \endinsert
}
\myfig {1} {0.4}
{\includegraphics{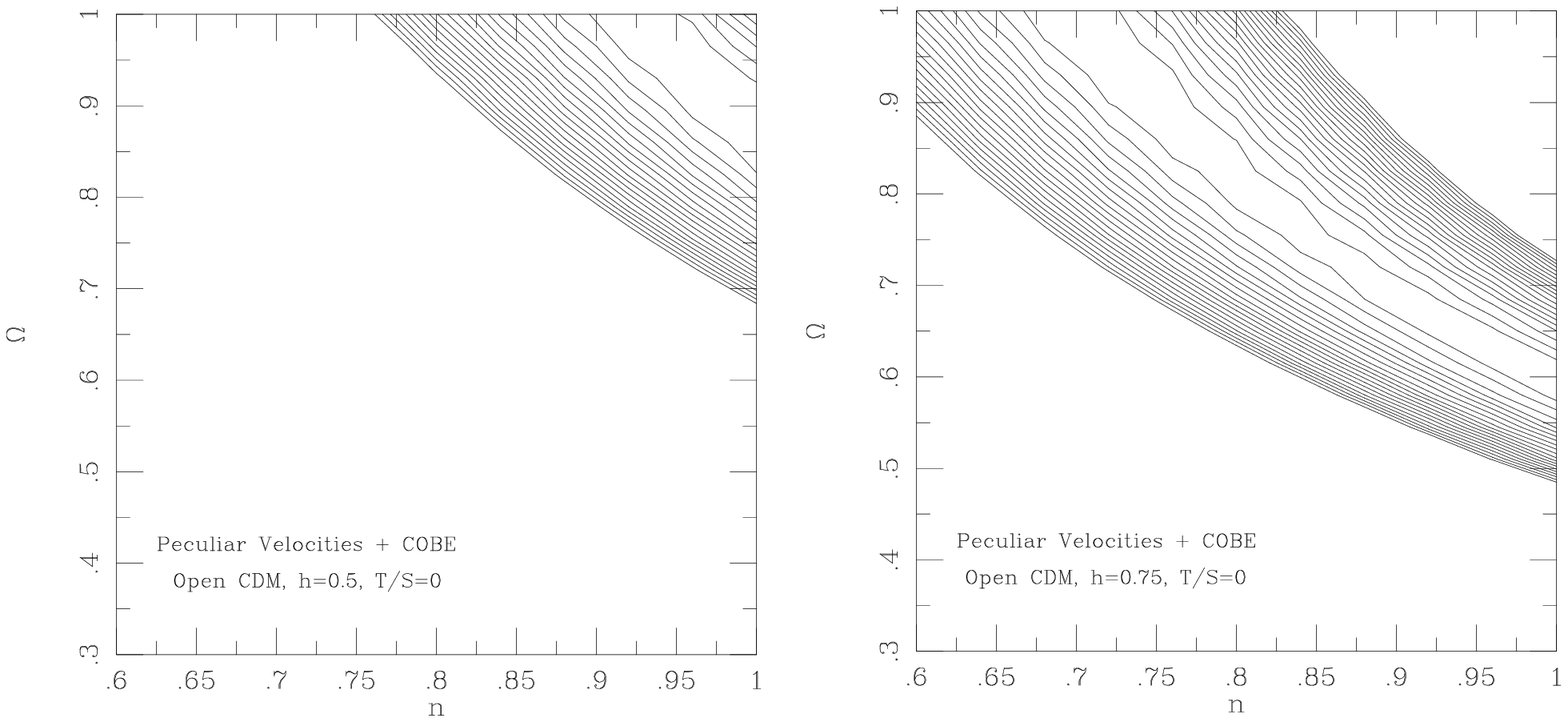}}
{
{\sbf Figure 5:} {\srm The same as Figure 4 for tilted Open CDM models without
tensor component, ${\ss \Delta[\ln \cL] = -1}$.}
\vskip 0.3 true cm
}

Figure 5 shows the analogous likelihood results in the $n-\Omega$ plane
for the tilted-\ocdm\ family of models (normalized by Sugiyama 1995).
Computations were done for the two values of $h$ as before,
but this time only for the case of no tensor fluctuations.
The results are qualitatively similar to the case of tilted-\lcdm\ models.
The tight constraint for tilted-\ocdm\ is
$$ 
\Omega\, {\h5}^{0.95}\, n^{1.4} = 0.88\pm 0.09\,.
\eqno(14)
$$
In this case, the trends along the ridge of high likelihood are weak.

\medskip\noindent\secpush\noindent
{\it 4.2.3 Model Dependence}

The most likely PS for the tilted-\lcdm\ family of models with tensor
fluctuations and $h=0.75$, corresponding to $\Omega=0.97$ and $n=0.84$,
is plotted in Figure 6.
The figure also shows the typical scatter in the PS about the most likely
model, using $n-\Omega$ pairs that lie on the innermost likelihood
contour (Fig.\ 4, lower-right panel).
This scatter is somewhat smaller than in the $\Gamma$ model (Fig.\ 2b),
partly due to the additional constraint from COBE.
Within this family of models, the constraints obtained are
$P(k)\Omega^{1.2}=(4.2\pm0.8)\times 10^3 \3hmpc$ at $k=0.1\ihmpc$
and $\sigma_8 \Omega^{0.6}=0.88 \pm 0.10$, where the error-bars are
the likelihood analysis $90\%$ confidence limits.
The scatter in these quantities for the other CDM families of
models is roughly the same.

\def\myfig#1#2#3#4{
  \midinsert
     \vskip #2 true cm \vskip 2.3 true cm
     #3
     \vskip 5.2 true cm
     {\baselineskip 11pt
       \rightskip=0.1 true cm \leftskip=0.1 true cm
       \vskip -0.3 true cm
       \par\noindent{\figtitle}\hskip -0.03 true cm #4 \par
       \rightskip=0 true cm \leftskip=0 true cm
     }
     \vskip -0.2 true cm
  \endinsert
}
\myfig {1} {0.4}
{\includegraphics{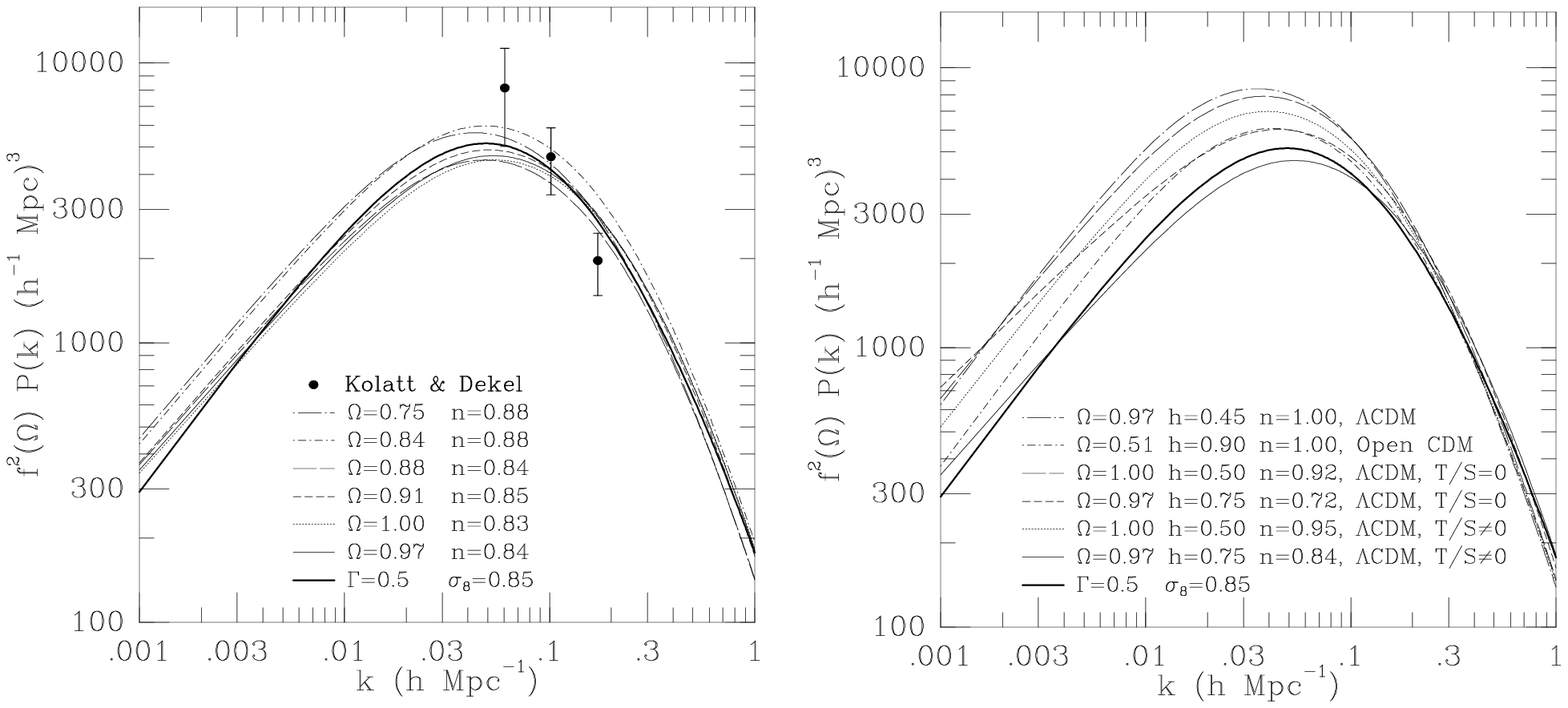}}
{{\sbf Figure 6:} {\srm
The PS of the most probable COBE-normalized CDM
model (solid), and the scatter about it following parameter pairs
that lie on the innermost likelihood contour (Fig. 4, lower-right panel).
The COBE-free ${\ss \Gamma}$ model is also marked (heavy solid).
The PS computed by KD from POTENT density of the same
velocity data (independent of COBE or models), 
and their ${\ss 1 \sigma}$ measurement errors, are shown in three bins.}
\vskip 0 true cm
\noindent
{\sbf Figure 7:} {\srm The best-fit power spectra for various CDM models.}
\vskip 0 true cm
}

How robust are the results to the choice of model within the CDM families?
\hbox{Table 1} in Appendix A shows the features of the most likely models
for each family of CDM variant discussed above, for the G catalog.
Listed for each model is the corresponding value of $\sigma_8 \Omega^{0.6}$,
the amplitude of the PS at $k=0.1\ihmpc$, the location of the peak
and, most important, the functional fit to the high-likelihood ridge, in
which the most likely model resides.

These power spectra of the best-fit models
are plotted in Figure 7.
All the models agree to within $\sim\! 20\%$ for $k>0.1\ihmpc$, and
they differ by up to $30-50\%$ on larger scales.
The amplitude of the PS at $k=0.1\ihmpc$ across the models
is constrained to be $P(k)\Omega^{1.2}=(4.8\pm0.8)\times 10^3 \3hmpc$.
The value of $\sigma_8 \Omega^{0.6}$ is roughly the same for all
these models and ranges between 0.87 and 0.94.

Table 1 shows as well the $\ln \cL$ value of each of these
models (with the zero set arbitrarily to the highest likelihood).
The best-fit models are all of comparable likelihood.
The most likely model, with a small margin,
is the tilted-\lcdm\ model with tensor fluctuations and $h=0.75$:
$\Omega=0.97$ and $n=0.84$, the model plotted in Figure 6.
This conclusion, however, is limited to the G catalog.
For the S catalog, the preferred models are non-tilted ($n=1$),
with lower $\Omega$ ($\sim 0.6-0.7$) and $h$ ($\sim 0.6$) (see
Table 2).

Figure 6 also compares the PS corresponding to the most-likely
COBE-normalized CDM model and
the most likely $\Gamma$ model that is independent of COBE.
The agreement between the two is within $\sim\! 10\%$ on all scales.
The agreement on the location of the peak in the PS reflects the fact 
that the peculiar velocity data themselves contain some meaningful    
information on scales as large as the size of the Mark III sample     
even without the constraint from COBE.                                

The PS computed by KD from the POTENT smoothed density field, that has
been recovered from the same Mark III data independent of COBE or
models, is also displayed in Figure 6.
The results (for all the models tested here) agree within 1$\sigma$ of the
measurement errors, and they agree particularly well near $k=0.1\ihmpc$,
where the velocity data imposes the strongest constraints.
The spectral slope $m$ near $k=0.1\ihmpc$, as constrained     
by our method, is roughly $m=-0.75 \pm 0.45$. The KD spectrum 
appears to be steeper ($m \sim -1.5$),                        
and the associated estimate of $\sigma_8 \Omega^{0.6}$ by KD is therefore
marginally lower, at $\simeq 0.7-0.8$.
In fact the KD slope is steeper than any of the CDM spectra discussed here
(see Figure 7), and is roughly as steep as the PS predicted by the CHDM
model, a 7:3 mixture of cold and hot dark matter (KD).
This may indicate that the CDM family, with the exception of CHDM,
does not allow enough freedom for a perfect fit on small scales.
Also, our present results may be less reliable on small
scales because, other than grouping, we do not make here any correction
for nonlinear effects.
On the other hand,
the result of the current paper is probably more reliable than KD on
large scales, because the likelihood method uses all the velocity data
including the large-scale flows,
while the POTENT density field is insensitive to the bulk velocity.

\def\myfig#1#2#3#4{
  \midinsert
     \vskip #2 true cm \vskip 2.3 true cm
     #3
     \vskip 5.3 true cm
     {\baselineskip 11pt
       \rightskip=0.1 true cm \leftskip=0.1 true cm
       \par\noindent{\figtitle}\hskip -0.03 true cm  #4 \par
       \rightskip=0 true cm \leftskip=0 true cm
     }
     \vskip -0.4 true cm
  \endinsert
}
\myfig {1} {0.4}
{\includegraphics{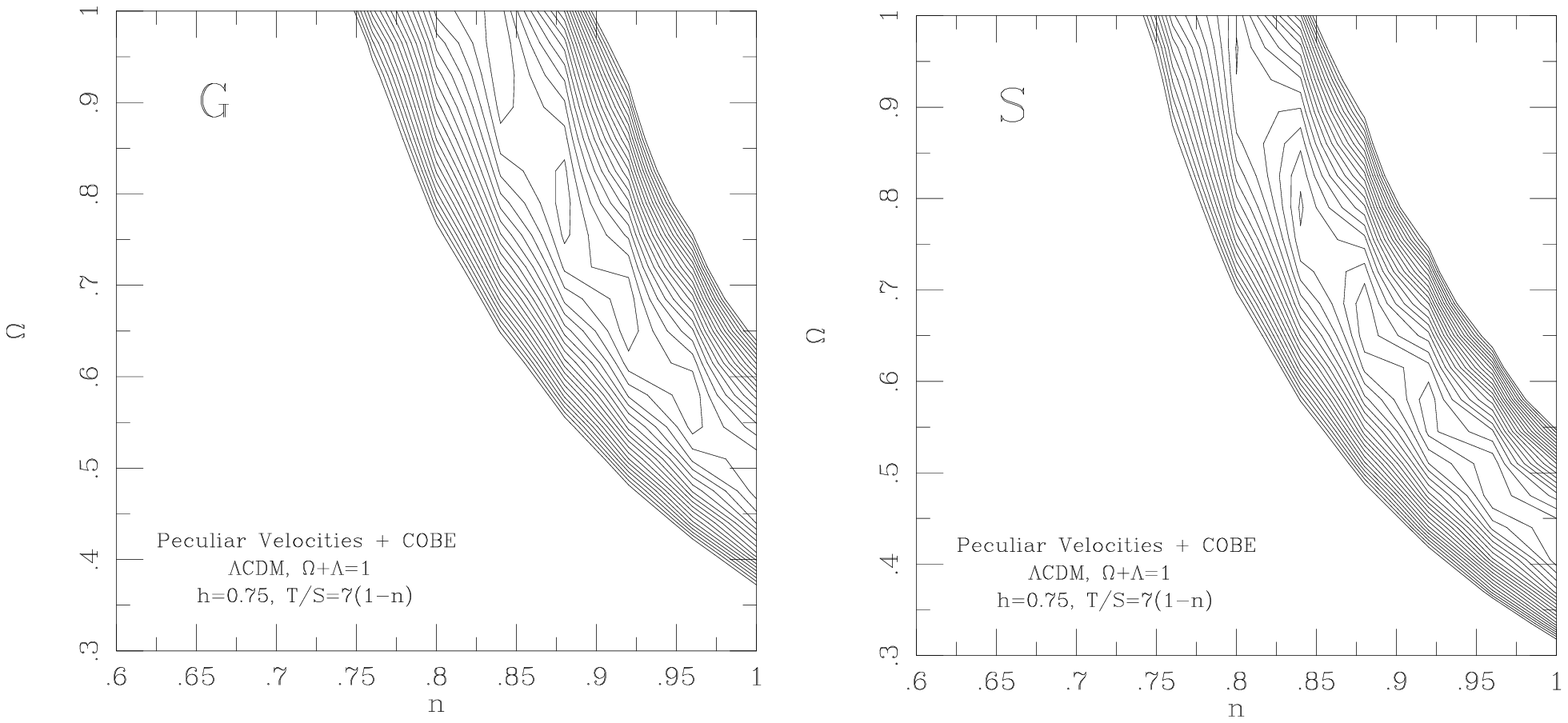}}
{
{\sbf Figure 8:} {\srm Contours of ${\ss \ln}$-likelihood in the
${\ss n-\Omega}$ plane, calculated for tilted-\sslcdm\ model with a
tensor component and ${\ss h=0.75}$,
obtained from the grouped data (G, same as Fig. 4, lower-right
panel) and from the singles data (S). ${\ss \Delta[\ln \cL] = -1}$.}
\vskip 0.3 true cm
}

\medskip
\cntl{\it 4.3. Robustness to Grouping in the Data}

To test robustness to the grouping and Malmquist correction in the
Mark III data, we have performed a similar likelihood analysis to
the S catalog.
Table 2 in Appendix A is the equivalent of Table 1 and summarizes
the results for the S catalog.
The S likelihood analysis might be more susceptible to
small-scale non-linear effects that are not properly dealt with in our method.
Still, it is interesting to compare the results for the G and S catalogs
in order to distinguish further between robust and non-robust conclusions.

Figure 8 shows, as an example, the results obtained from the G and S catalogs
for the tilted-\lcdm\ model with tensor fluctuations and $h=0.75$.
The resemblance between these results is typical of the
other models as well. First, the high-likelihood ridges are similar in shape:
the powers $\mu$ and $\nu$ are almost the same in the G and
S cases for all the CDM models we checked. Second, there is a slight
difference in the constant defining the ridge: for the model plotted here
the S result is $\Omega\, {\h5}^{1.25}\, n^{3.4} = 0.71 \pm 0.12$
while the G ridge was $\Omega\, {\h5}^{1.3}\, n^{3.4} = 0.85 \pm 0.12$,
preferring somewhat higher values for the parameters.
A similar trend occurs for all the models (see Appendix A).

The gradient of likelihood along the ridge is different for the G and S
catalogs to the extent that it is sometimes of opposite sign.
The G data slightly prefers a tilt in
the spectrum and $\Omega \simeq 1$,
while the S data slightly favors lowering the value of $\Omega$
below unity and $n \simeq 1$.
This difference indicates that we should not assign high significance to the
trends along the ridges of high likelihood and rather
focus on the robust ridges themselves.
We thus quote constraints on the degenerate combinations of $\Omega$,
$h$ and $n$, and not on any of them separately.

The likelihood analysis of the S data with the $\Gamma$ model
also yields some differences. The maximum likelihood value
is $\Gamma=0.3 \pm 0.1$ for the S catalog, compared to the
$\Gamma=0.5 \pm 0.15$ for the G catalog. The shape parameter
$\Gamma$ (independent of COBE normalization) seems to be only
poorly constrained by the data. Still, we can quote based on
the velocity data alone that $\Gamma \ge 0.2$ at the $90\%$ confidence level.

The robust feature for the $\Gamma$ model is the value of
$\sigma_8 \Omega^{0.6}$,
which is stable at $\sigma_8 \Omega^{0.6} = 0.85 \pm 0.1$ for both
the G and S catalogs. The lower $\Gamma$ from the S
data is accompanied by a higher amplitude $A$, such that the
combination $\sigma_8 \Omega^{0.6}$ remains unchanged.

Another robust feature, for all models, is the amplitude of the
maximum-likelihood power spectrum on intermediate scales.
The PS amplitude at $k=0.1\ihmpc$ for both S and G is constrained
to be $P(k) \Omega^{1.2} = (4.8\pm1.5)\times 10^3 \3hmpc$.
The location of the peak in the PS in the two cases is roughly the same,
ranging between $0.025$ and $0.04 \ihmpc$ for the S catalog, and
between $0.03$ and $0.06 \ihmpc$ for the G catalog.

\medskip\bigskip\secpush
\cntl{\bf 5. CONCLUSION}

We have presented a Bayesian method for deriving the power
spectrum of mass density fluctuations
from the \m3 Catalog of Peculiar Velocities.
The result is free of galaxy ``biasing."
The method extracts the maximum amount of useful information from the data.
It is exact to first order, under the assumption of Gaussian
fluctuations and Gaussian errors.
Tests using realistic mock catalogs show that
this approximation is adequate because of the large-scale coherence of
the velocity field and because the large errors that dominate on
small scales make the nonlinear effects contribute only
weakly to the result.
The formal likelihood errors include both measurement errors        
and cosmic scatter.                                                 
Ignored are the errors in the COBE normalization, of order of a few 
percent.                                                            

Our robust result for the whole family of models examined here as priors
and for the different ways of handling the grouping of the data is
that the mass PS amplitude at $k=0.1\ihmpc$ is
$P(k) \Omega^{1.2} = (4.8\pm1.5)\times 10^3 \3hmpc$.
The robust integral constraint for the CDM family of models  
$\sigma_8 \Omega^{0.6}=0.88\pm 0.15$. 
The errors quoted are crude: they reflect the typical 90\% uncertainty
for each of the best-fits within each family of models, combined with
the typical scatter among these best-fit models and between the G and S
treatments of the data.
Similar results are obtained when using the velocity data alone and when
the additional constraints from COBE are included.
For the span of models checked, the PS peak is in the range
$0.02\leq k\leq 0.06\ihmpc$.  

This normalization of the PS is in pleasant agreement with the independent
`frequentist' computation of the PS via POTENT by KD, which yielded
$P(k) \Omega^{1.2} = (4.6\pm1.4)\times 10^3 \3hmpc$ at $k=0.1\ihmpc$,
and $\sigma_8 \Omega^{0.6}\approx 0.7-0.8$.
Our new results differ at about the $2\sigma$
level from the earlier estimate by Seljak \& Bertschinger (1994)
of $\sigma_8 \Omega^{0.6}=1.3 \pm 0.3$.
They performed a likelihood analysis of the POTENT reconstructed
density field based on the earlier Mark II sample.
The main improvements since then are that the current
analysis includes five times denser sampling in a more extended volume,
the systematic errors such as Malmquist bias are handled better,
and a wider span of models and parameters
is used in the likelihood analysis.
It may be interesting to note that
the current measurement is somewhat higher than the completely
independent estimate of a similar quantity
based on cluster abundances, $\sigma_8 \Omega^{0.56} \simeq 0.57\pm0.05$
(White, Efstathiou, \& Frenk  1993), but it is only about $2\sigma$ away.

The comparisons of the mass $\sigma_8$ to the values observed for optical
galaxies ($\simeq 0.95$) and for IRAS 1.2Jy galaxies ($\simeq 0.6-0.7$)
indicate $\beta$ values
of order unity to within 25\% for most galaxy types on these scales
(see KD, Fig. 6 and Table 2, for more details).

A $\Gamma$-shape model, free of COBE normalization, is constrained by the
velocity data only weakly, to $\Gamma$ in the range $0.2-0.6$.
The most likely value of $\Gamma$ are
somewhat higher than the canonical values of $\sim 0.2-0.3$ typically
obtained from galaxy density surveys (\eg Efstathiou \etal 1992, Peacock \&
Dodds 1994).

Within the families of COBE-normalized CDM models,
which we have restricted to the range $\Omega\leq 1$ and $n\leq 1$,
we have obtained constraints on combinations of the cosmological
parameters of the sort $\Omega\, {\h5}^\mu\, n^\nu = 0.8 \pm 0.2$, with
$\mu=1.3$ for the flat models and with $\nu=3.4$ and $2.0$
with and without tensor fluctuations respectively.
For the open models without tensor fluctuations the corresponding
powers are $\mu=0.95$ and $\nu=1.4$.
The extended error quoted include 
the uncertainty due to measurement errors and finite sampling, 
the variation among the various CDM models, and the uncertainty in 
the best way to group the data for Malmquist correction. 

The most likely model for the G data is a tilted-\lcdm\ model with tensor
fluctuations, relatively high Hubble constant ($h \sim 0.75$), $\Omega$
near unity and $n \sim 0.85$.
These results are consistent with the conclusion of
White \etal (1995), who argue for tilted CDM based on several data
sets including power spectra of galaxy density
(Peacock \& Dodds 1994), cluster correlations, pair-wise velocities
and COBE's results. Their best fit is
$\Omega=1$, $h\approx 0.45$, $n=0.9$, with tensor fluctuations.
This model is about 1$ \sigma$ away from our best fit
but the PS is quite similar.
Our lower value of $n$ compensates for the higher value of $h$.

It is interesting to note that based on the velocity data
and COBE normalization alone, the standard CDM model
($\Omega=1$, $n=1$, $h=0.5$) is
less likely than its various variants studied here (see Appendix A).
A slight tilt in $n$ or a small decrease of $\Omega$ can increase
the likelihood significantly. However, this does not imply that the
standard CDM model is necessarily ruled out as our analysis measures
only relative likelihoods rather than absolute goodness of fit.

In a subsequent paper (Zaroubi \etal 1996), our range of allowed power
spectra from the peculiar velocities is translated to an angular power
spectrum of CMB fluctuations and compared to recent subdegree observations.
We find there that in order to also fit the height of the first CMB peak with a
CDM spectrum, the tilt cannot be more pronounced than $n \sim 0.9$, 
the total energy density $\Omega+\Lambda$ should be close to unity,
and the baryonic fraction should be large, $\Omega_b \sim 0.1$. A
discussion of our results in the wider context of cosmological
parameters can be found in Dekel, Burstein \& White (1997) and Dekel (1997).

It is worth recalling that the recovered power spectrum is highly sensitive
to the assumed observational errors, because the noise has a systematic effect
on the measured power spectrum. Effectively, the power spectrum
of the errors has to be subtracted out. This can be done explicitly,
as in KD, or implicitly, as in our current likelihood analysis. Our
input error estimate is based on the careful error analysis of the
Mark III data by Willick \etal (WI; WII; WIII). If, for some reason,
this is an underestimate of the errors, then the recovered power
spectrum is an overestimate, and vice versa.

Finally, a note of caution of technical nature about the method
and its application to the current data under certain extreme conditions.
If not enough constraints are imposed,
the inversion of $\tilde U_{i,j}$ (Eq.~5) may become dominated
by the noise rather than the signal and might lead to a wrong answer.
We encountered this problem when we tried to parameterize the PS
with a multi-parameter function that did not enforce
any upper bound on the power on large scales
(bounds that are properly enforced when COBE constraints are used, for
example).
The likelihood analysis preferred in this case
unphysically high power on large scales. This is probably, at least in part,
a result of noise dominance, and is beyond the scope of this paper.
An algorithm to detect and possibly eliminate this problem
is discussed elsewhere (Zaroubi 1995).
As long as we use models that are properly bound on large scales
the results on small scales are robust.
The $\Gamma$ model, for example, does not suffer from this problem because
its shape effectively enforces the appropriate bounds on large scales.
The data strongly constrain the amplitude $P_{0.1}$ and the slope 
near $k=0.1\ihmpc$. 
A $\Gamma$ model with the peak at a very small $k$ and the same $P_{0.1}$
would have required a steeper slope ($m=-2$ to $-3$)
which is disfavored by the data.
The $\Gamma$ model thus enforces specific connections between
properties that are strongly
constrained by the data with properties that are poorly
constrained.

\bigskip\secpush
\cntl{\bf Acknowledgments}

We thank Naoshi Sugiyama for providing the COBE normalizations for the CDM
models prior to publication and for close interaction.
We acknowledge stimulating discussions with
Marc Davis, Ravi Sheth, Douglas Scott and Martin White.
We thank the referee Robert Crittenden for helpful comments. 
YH and IZ acknowledge the hospitality of the Astronomy Department
and the Center for Particle Astrophysics at Berkeley where part of
this work was done.
This work is supported in part by the US-Israel Binational Science Foundation
grants 92-00355, 95-00330 and 94-00185, by the Israel Science
Foundation grants 469/92, 950/95 and 590/94, and by the US National
Science Foundation grant PHY-91-06678.
IZ has been partly supported by the Exchange of Astronomers Programme
of the IAU.
\bigskip\bigskip
\vskip 0.5 true cm

\cntl{\bf APPENDIX A}

We present here the detailed likelihood analysis results for the various
COBE normalized CDM families of models. Table 1 shows the results obtained for
the G catalog. The functional fit of the high-likelihood ridge obtained
in each case is given. The error-bar quoted is the formal $90\%$
confidence limit of
the degenerate parameter combination. The values of the cosmological
parameters corresponding to the maximum-likelihood model within this
ridge are listed as well. Parameters that were held fixed in the
likelihood analysis while varying the other parameters are denoted by
``fixed''. For reference, the
COBE normalized standard CDM model is also listed.
The value of $\ln {\cal L}$  obtained for each of these models is given,
with the zero set arbitrarily to correspond to the most likely model.

\def\myfig#1#2#3#4{
  \midinsert
     \vskip #2 true cm \vskip 2.3 true cm
     #3
     \vskip 7.3 true cm
     {\baselineskip 11pt
       \rightskip=0.1 true cm \leftskip=0.1 true cm
       \vskip -0.3 true cm
       \par\noindent{\figtitle}\hskip -0.03 true cm #4 \par
       \rightskip=0 true cm \leftskip=0 true cm
     }
     \vskip -1.0 true cm
  \endinsert
}

\myfig {1} {0.4}
{\includegraphics{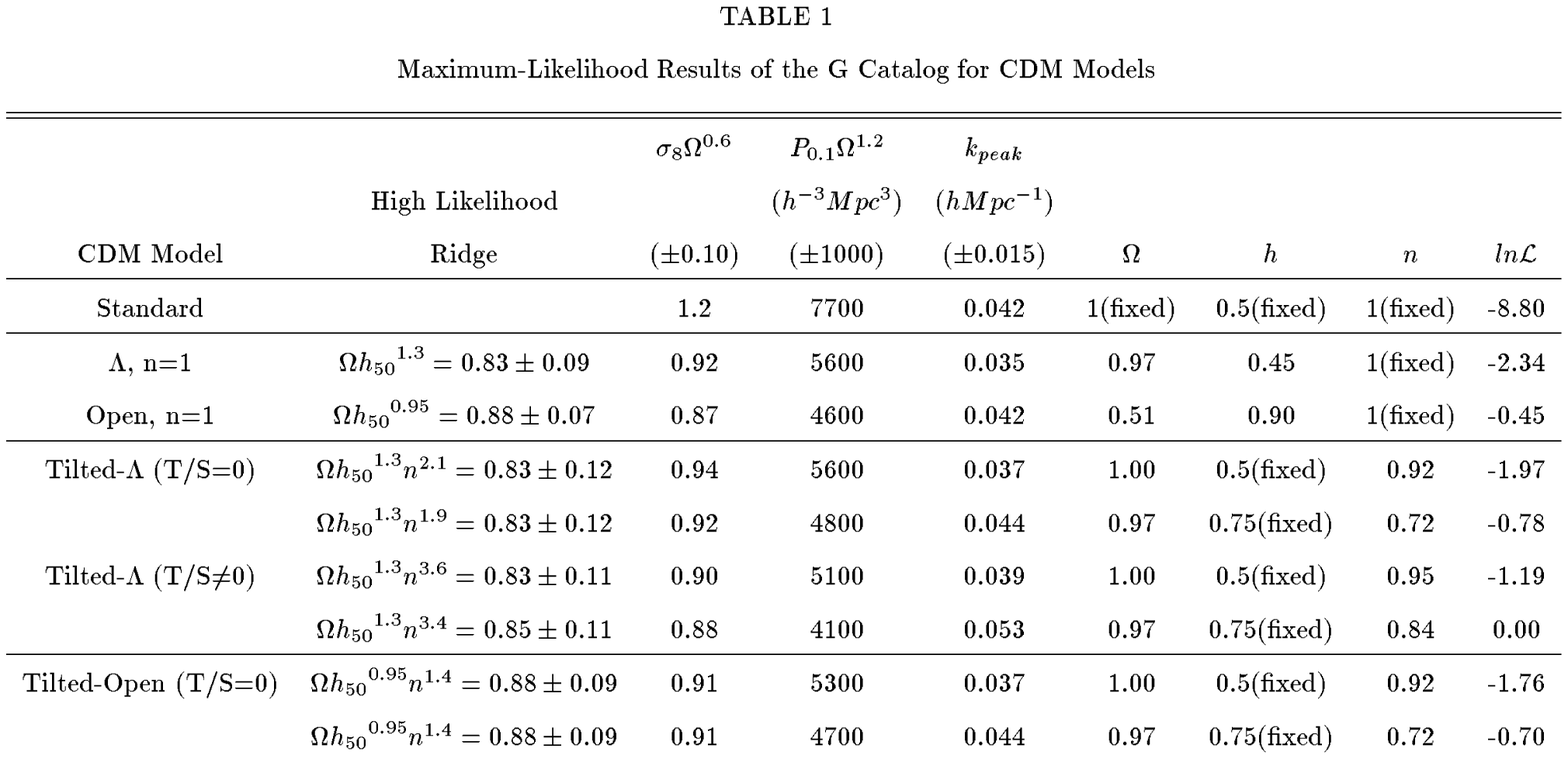}}
{}

Several features of the power spectra of these best-fitting models
are  presented: the
value of $\sigma_8 \Omega^{0.6}$, the amplitude of the PS at
$k=0.1\ihmpc$, $P_{0.1} \Omega^{1.2}$, and the location of the peak in
the PS, $k_{peak}$. The quoted error-bars in the table headings
represent the typical $90\%$ likelihood
uncertainty in these quantities within each family of models.
Due to the high-likelihood ridge in each family of models,
the exact location of the maximum-likelihood model within the ridge is
hardly significant, but the table demonstrates the robust features
among these models (see discussion in \S 4.2.3).

Table 2 lists the results of the various CDM models for the S catalog.
Similar high-likelihood ridges are listed, though there is a small
systematic shift toward smaller values in the constant defining the
ridge compared to the G data.
The trends along the ridges and the likelihood ranking of the various
models are different then in the G case but, again, they are only
marginally significant.  However the general features
of the PS, such as the values of $\sigma_8 \Omega^{0.6}$ and
$P_{0.1} \Omega^{1.2}$, are similar (see \S 4.3).

\myfig {1} {0.4}
{\includegraphics{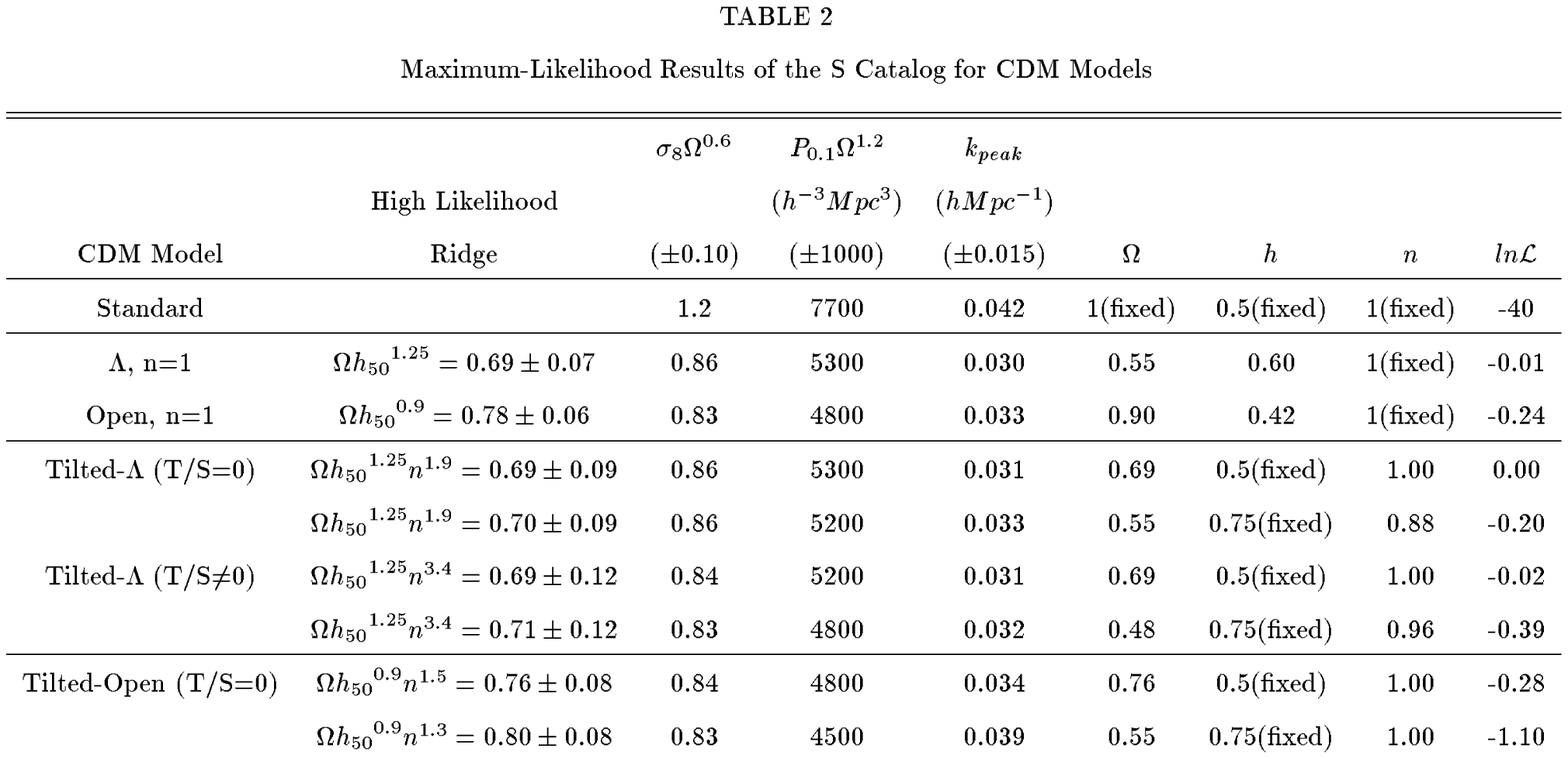}}
{\vskip 0.5 true cm}

\bigskip\bigskip
\cntl{\bf REFERENCES}

\prref Bardeen, J. M., Bond, J. R., Kaiser, N., \& Szalay, A. S. 1986, \ApJ,
304, 15

\prref Bennett, C. L., \etal 1996, \ApJ, 464, L1

\prref Bertschinger, E., \& Dekel, A. 1989, \ApJ, 336, L5

\prref Crittenden, R., Bond, J. R., Davis, R. L., Efstathiou G., \& 
Steinhardt, P. J. 1993, Phys. Rev. Lett, 61, 324

\prref Dekel, A. 1994, \ARAA, 32, 371

\prref Dekel, A., 1997, in Formation of Structure in the Universe, ed.
A. Dekel \& J. P. Ostriker (Cambridge: Cambridge University Press), in press

\prref Dekel, A., Bertschinger, E., \& Faber, S. M. 1990, \ApJ, 364, 349

\prref Dekel, A., Burstein, D., \& White, S. D. M. 1997, in Critical
Dialogues in Cosmology, ed. N. Turok (Princeton: Princeton University
Press), in press

\prref Dekel, A., \& Rees, M. J. 1987, Nature, 326, 455

\prref Dressler, A. 1980, \ApJ, 236, 351

\prref Efstathiou, G., Bond, J. R., \& White, S. D. M. 1992, \MNRAS, 258, 1p

\prref G\'orski, K. M. 1988, \ApJ, 332, L7

\prref G\'orski, K. M., Ratra, B., Sugiyama, N., \& Banday, A. J. 1995,
\ApJ, 444, L65

\prref Jaffe, A. H., \& Kaiser, N. 1995, \ApJ, 455, 26 

\prref Kaiser, N. 1987, \MNRAS, 227, 1

\prref Kaiser, N. 1988, \MNRAS, 231, 149

\prref Kolatt, T., \& Dekel, A. 1996, \ApJ, in press 
(astro-ph/9512132) 

\prref Kolatt, T., Dekel, A., Ganon, G., \& Willick, J. 1996, \ApJ, 458, 419

\prref Peacock, J. A., \& Dodds, S. J. 1994, \MNRAS, 267, 1020

\prref Press, W. H., Teukolsky, S. A., Vetterling, W. T., \& Flannery,
B. P. 1992, "Numerical Recipes'' (2d ed.; Cambridge: Cambridge
University Press)

\prref Seljak, U., \& Bertschinger, E. 1994, \ApJ, 427, 523

\prref Sugiyama, N. 1995, \ApJSup, 100, 281

\prref Turner, M. S. 1993, Phys. Rev., D48, 5302

\prref Tytler, D., Fan, X-M., \&  Burles, S. 1994, \Nature, 381, 207

\prref White, M., \& Bunn, E. F. 1995, \ApJ, 450, 477

\prref White, M., Scott, D., Silk, J., \& Davis, M. 1995, \MNRAS, 276, 69P

\prref White, S. D. M., Efstathiou, G., \& Frenk, C. S. 1993, \MNRAS, 262, 1023

\prref Willick, J. A., Courteau, S., Faber, S. M., Burstein, D., \&
Dekel, A. 1995, \ApJ, 446, 12

\prref Willick, J. A., Courteau, S., Faber, S. M., Burstein D., Dekel, A., \&
Kolatt, T. 1996a, \ApJ, 457, 460

\prref Willick, J. A., Courteau, S., Faber, S. M., Burstein, D., \&
Dekel, A. 1996b, \ApJSup, in preparation

\prref Zaroubi, S. 1995,
in Proc. of the XXXth Rencontres de Moriond ``Clustering in the
Universe'', ed. S. Maurogordato, C. Balkowski, C. Tao \& J. Tr\^an Thanh
V\^an (Gif-sur-Yvette Cedex: Editions Frontieres), 135 

\prref Zaroubi, S., \& Hoffman, Y. 1996, \ApJ, 462, 25

\prref Zaroubi, S., Hoffman, Y., \& Dekel, A. 1996, preprint

\prref Zaroubi, S., Hoffman, Y., Fisher, K. B., \& Lahav, O. 1995,
\ApJ, 449, 446

\prref Zaroubi, S., Sugiyama, N., Silk, J., Hoffman, Y., \& Dekel, A. 
1996, \ApJ, submitted 
(astro-ph/9610132) 

\bye